\documentclass[aps,preprint,prd,floatfix,nofootinbib]{revtex4-1}%
\usepackage{amsmath, amsfonts}
\usepackage{bm}
\usepackage{color}
\usepackage{amsmath}
\usepackage{amsfonts}
\usepackage{amssymb}
\usepackage{graphicx}%
\usepackage[utf8]{inputenc}
\usepackage{amsfonts}%
\usepackage{amsmath}%
\usepackage{amssymb}%
\usepackage{graphicx}
\providecommand{\U}[1]{\protect\rule{.1in}{.1in}}

\begin{document}
\title{Superradiance of charged black holes in Einstein-Gauss-Bonnet Gravity}
\author{Octavio Fierro}
\email{ofierro@ucsc.cl}
\affiliation{Departamento de Matem\'{a}tica y F\'{\i}sica Aplicadas, Universidad
Cat\'{o}lica de la Sant\'{\i}sima Concepci\'{o}n, Alonso de Rivera 2850,
Concepci\'{o}n, Chile.}
\author{Nicol\'{a}s Grandi}
\email{grandi@fisica.unlp.edu.ar}
\affiliation{Instituto de F\'{\i}sica de La Plata - CONICET  \&  Departamento de F\'{\i}\i
sica - UNLP,\\ C.C. 67, 1900 La Plata, Argentina}
\author{Julio Oliva}
\email{juoliva@udec.cl}
\affiliation{Departamento de F\'{\i}sica, Universidad de Concepci\'{o}n,
\\ Casilla, 160-C,
Concepci\'{o}n, Chile}

\begin{abstract}
In this paper we show that electrically charged black holes in Einstein-Gauss-Bonnet gravity suffer from a superradiant instability. It is triggered by a charged scalar field that fulfils Dirichlet boundary conditions at a mirror located outside the horizon. As in General Relativity, the unstable modes exist provided the mirror is located beyond a critical radius, making the instability a long wavelength one. We explore the effects of the Gauss-Bonnet corrections on the critical radius and find evidence that the critical radius decreases as the Gauss-Bonnet coupling $\alpha$ increases. Due to the, up to date, lack of an analytic rotating solution for Einstein-Gauss-Bonnet theory, this is the first example of a superradiant instability in the presence of higher curvature terms in the action.

\end{abstract}
\maketitle

\section{Introduction}

Classically, black hole horizons effectively work as a one way membrane that
precludes information to be transferred from the inner to the outer region. In
spite of this defining feature, energy can be extracted from rotating black
holes by superradiant scattering, in a process in which an impinging wave gets
amplified by the interaction with the black hole (see the seminal works \cite{uno}-\cite{tres} as well as a more recent thorough review \cite{Brito:2015oca}). Given the fact that the
black hole entropy can only increase in such process, from the first law of
black hole thermodynamics it can be shown that the energy extraction can occur
only if the black hole possesses a non-vanishing global charge \cite{bekenstein}, on top of the mass. In the case of rotating black holes such a charge is provided by the angular momentum.
It can be shown that the superradiant scattering
takes place provided $\omega_R <{ m}\Omega_{h}$,
where $\omega_R$ is the real part of the scalar field frequency, ${ m}$ stands for its ``magnetic quantum
number" and $\Omega_{h}$ is the angular velocity of the horizon.

During the last decades gravity in higher dimensions has received considerable
attention \cite{Horowitz:2012nnc}. Either in its own right
or motivated by the task of understanding the low-energy limit of string
theory, gravity in dimensions greater than four has permitted to explore how
generic are the features that black holes have in four dimensions. In this scenario, higher
curvature corrections naturally appear, as $\alpha^{\prime}$ corrections or
as an answer to the natural question of which is the most general theory that
preserves diffeomorphism invariance and has second order field equations. In
the context of the latter, a whole new set of theories appear in dimensions
greater than four: the Lovelock theories \cite{Lovelock:1971yv}. The Lagrangian densities
of these theories are defined by dimensional continuations of the even
dimensional Euler densities, exactly in the same manner than the
Einstein-Hilbert Lagrangian is the dimensional continuation of the Euler
density for Euclidean, compact manifolds, without boundary in dimension two.
The simplest Lovelock theory that departs from general relativity is the
Einstein-Gauss-Bonnet theory. The theory supplements the
Einstein-Hilbert action with a term that is quadratic in the curvature and
contains a new dimensionfull constant, $\alpha$, of mass dimension $-2$.
This theory can be obtained from $\alpha^{\prime}$ corrections in string
theory \cite{Zwiebach:1985uq}. For generic values of the constant $\alpha$, the
spherically symmetric solution of Einstein-Gauss-Bonnet \cite{Boulware:1985wk} turns out to be static
\cite{Zegers:2005vx}-\cite{Deser:2005gr} and the lapse function is determined by a quadratic equation,
actually giving rise to two solutions. Only one of these solutions leads to
Schwarzschild-Tangherlini when $\alpha\rightarrow0$, and is therefore well defined as a correction of the latter. Such solution is known as
the Einstein branch.

A natural question therefore arises: what's the effect of the quadratic
correction on the superradiant phenomenon? In order to answer that question
one would first need a rotating solution for the Einstein-Gauss-Bonnet theory. Such solution has
not been found in an analytic and closed form making impossible to study the
superradiant instability in Einstein-Gauss-Bonnet in the same manner than one does in GR.
Efforts for constructing rotating solutions on Einstein-Gauss-Bonnet had rely on numerical
\cite{Brihaye:2008kh} or perturbative tools \cite{Kim:2007iw}. Including a cosmological term
in the action, it was shown that in five dimensions and for a precise relation
between the couplings, a rotating solution within the Kerr-Schild family can
be constructed \cite{Anabalon:2009kq}. Whether or not the latter non-circular metric
represents a black hole is not known \cite{Anabalon:2010ns}.

As stated in the original reference \cite{bekenstein}, the electric charge may also provide for a global charge that could allow for a superradiant process whenever charged fields are involved. Indeed, charged black holes also suffer from a superradiant scattering, and due to the lack of an analytic rotating solution in the Einstein-Gauss-Bonnet theory, they offer
a natural arena to study the superradiant energy extraction from a black hole
in the presence of quadratic terms. For this charged case, in order to access
a superradiant regime in general relativity in four dimensions, one needs a
charged scalar that can be confined in a region between the event horizon and
a reflecting mirror (see e.g. \cite{Hod:2013eea} and \cite{Degollado:2013eqa}). For a given mass and charge of the black hole hole, as well as mass and charge of the
scalar, the mirror has to be located above a certain minimum, critical, radial
coordinate distance from the horizon. It has been recently shown that the
maximum time grow for the field in the charged case exceeds the corresponding
one of the rotating case by a factor of almost $10^{3}$ \cite{Herdeiro:2013pia}.
This was particularly useful for the study of the final stage of the charged
superradiant instability, which leads to a charged hairy black hole
\cite{Sanchis-Gual:2016tcm}. Since for large values of the charge of the scalar this
turns out to be a highly energetic process, it's natural to expect that in
higher dimensions and in the presence of a quadratic term in the action, the
phenomenology could be modified.

\bigskip

As a first step, in this work we focus on the propagation of
a charged massive scalar on a charged black hole of the Einstein-Gauss-Bonnet theory. In Section II, we set
the problem by introducing the theory and its charged black hole solution in $D$ dimensions. In Section III we introduce the scalar field perturbation in the problem and discuss its superradiant scattering qualitatively. Then, in Section IV we show numerically that the superradiant scattering of a charged scalar in Einstein-Gauss-Bonnet indeed exists and that the minimum critical radius of the
mirror decreases as the Gauss-Bonnet coupling, $\alpha$, increases. We also provide a physical interpretation for this relation. For
simplicity we focus on dimensions five, six and seven. Section V contains some conclusions and
final remarks.

\section{Einstein-Gauss-Bonnet theory and its charged black holes}

The action for Einstein-Gauss-Bonnet theory in the presence of a Maxwell field and
a charged scalar $\phi$, reads
\begin{equation}
I\left[  g_{\mu\nu},A_{\mu},\phi\right]  =I_{grav}\left[  g_{\mu\nu}\right]
+I_{mat}\left[  A_{\mu},\phi,g_{\mu\nu}\right]  \ ,
\label{fullaction}
\end{equation}
where the gravitational part is given by
\begin{equation}
I_{grav}\left[  g_{\mu\nu}\right]  =\frac{1}{16\pi G}\int d^{D}x\sqrt
{-g}\left[  R+\frac{2\alpha}{(D \!-\!3)(D\!-\!4)}\left(  R^{2}-4R_{\mu\nu}R^{\mu\nu}+R_{\alpha\beta
\gamma\delta}R^{\alpha\beta\gamma\delta}\right)  \right]  \ ,
\end{equation}
and the matter part reads
\begin{equation}
I_{matt}\left[  A_{\mu},\phi,g_{\mu\nu}\right]  =\int d^{D}x\sqrt{-g}\left(
-\frac{(D\!-\!2)}{32\pi G(D\!-\!3)}F_{\mu\nu}F^{\mu\nu}-|D_{\mu}\phi|^{2}-\mu^{2}|\phi|^{2}\right)  \ ,
\end{equation}
with $D_{\mu}=\nabla_{\mu}-iqA_{\mu}$.
Here the Gauss-Bonnet coupling
$\alpha$ has mass dimension $-2$. As shown below, the factors we have included in front of the Gauss-Bonnet and Maxwell terms are useful for writing the charged black hole solution in a simple manner.

The theory (\ref{fullaction}) admits the following spherically symmetric black hole solution
\begin{equation}
ds^{2}=-f\left(  r\right)  dt^{2}+\frac{dr^{2}}{f\left(  r\right)  }%
+r^{2}d\Omega_{D-2}^{2}\ ,
\end{equation}
where $d\Omega_{D-2}$ is the line element of the $D-2$ sphere, and the lapse function $f(r)$ is given by
\begin{equation}\label{elf}
f(r)=1 +\frac{r^2}{4 \alpha} \left(1-\sqrt{1+\alpha  \left(
\frac{16 M}{ r^{D-1}}-\frac{8 Q^2} {r^{2 (D-2)}}\right)}\right)\,.
\end{equation}
This solution is supported by an electric potential given by
\begin{equation}
A=A_{t}\left(  r\right)  dt=-\frac{Q}{r^{D-3}}dt\ 
\ ,
\label{backmatter}
\end{equation}
and the scalar field vanishes $\phi=0$ \cite{wiltshire}. The integration constants $M$ and $Q$, relate to the mass and charge respectively. In general there are two possible signs in front of the square root in equation (\ref{elf}), but only the branch presented here leads to the solution of GR when the limit $\alpha\rightarrow 0$ is taken. See e.g. \cite{causals} for a thorough discussion of the allowed causal structures in the presence of a cosmological constant and as well as some thermal properties of the solutions.
\section{Scalar field perturbations}

As usual, since the energy-momentum tensor is at least quadratic on the matter
fields, the system that is obtained by turning on a scalar field perturbation
\[
\phi\rightarrow\phi+\psi\ ,
\]
is consistent to lowest order in $\psi$ and leads uniquely to the charged Klein-Gordon equation for
$\psi$ in the presence of the background solution defined by (\ref{backmatter})
\begin{equation}
\left(
\left(  \nabla-iqA\right)^2-\mu^2\right)
\psi=0\ .\label{eq}%
\end{equation}
In this equation, $\psi$ can be Fourier decomposed
in time and written as a superposition of hyperspherical harmonics
\begin{equation}
\psi\left(  t,r,\Omega\right)  =\sum_{lm_j}\int d\omega\ e^{-i\omega
t}R_{\omega l m_j}\left(  r\right)  Y_{l m_j}\left(  \Omega\right)
\ .\label{sep}
\end{equation}
Here $\omega=\omega_R+i\omega_I$ has in principle a real and imaginary part.
For simplicity hereafter we use the notation $R_{\omega l m_j}\left(  r\right)
=:R\left(  r\right)$. Here $l, m_j$ denote collectively the $D-2$ integers required to uniquely determine a $(D\!-\!2)$-hyperspherical harmonic which fulfil \cite{Higuchi:1986wu}
\begin{equation}
\nabla_{S^{D-2}}^{2}Y_{lm_j}\left(  \Omega\right)  =-l\left(  l+D-3\right)
Y_{lm_j}\left(  \Omega\right)  \ .
\end{equation}
Since the background is spherically symmetric, there is no ``magnetic splitting"
and only the eigenvalue $l$ appears on the equations. Finally, the equation
(\ref{eq}) for the radial dependence $R\left(  r\right) $ leads to%
\begin{equation}
\frac1{r^{D\!-\!2}}\left(
r^{D\!-\!2}f R^{\prime}
\right)^{\prime}
+
\left(  \frac{\left(\omega+qA\right)^2}{f}-\frac{l\left(
l\!+\!D\!-\!3\right)}{r^2} -\mu^{2}\right)  R=0\ ,\label{eqRarbD}%
\end{equation}
This equation has to be integrated numerically to obtain the radial profile. In order to determine the correct boundary condition at the horizon, we expand the equation in its vicinity, obtaining
\begin{equation}
4\pi T\left((r-r_+)
 R^{\prime}
\right)^{\prime}
+
\left(
\frac{\left(\omega-\omega_s\right)^2}{4\pi T (r-r_+)}
-\mu_l^{2}\right)
R=0\ .\label{eqRarbDapp}%
\end{equation}
where we defined $\mu_l^2=\mu^2+{l( l\!+\!D\!-\!3)}/{r_+^2}$ as an effective mass, $\omega_s=-qA_t(r_{+})$ as the largest superradiant frequency, and $T=f^{\prime}\left(  r_{+}\right)  /4\pi$ as the (non-vanishing) black hole temperature.
It is solved by
\begin{eqnarray}
R\left(  r\right) &=&
J_{\pm i \frac{\omega\!-\!\omega_s}{4\pi T}}\!
\left(\!\frac{\mu_l\sqrt{r-r_+}}{\sqrt T}\right)
\label{eq:bc1}\\
&=&
\left(  r-r_{+}\right)  ^{\pm i\frac{\omega-\omega_s }{4\pi T}}\left(1^{\phantom{2}}\!\!+\mathcal{O}(r-r_+)\right) \qquad \text{as }\qquad r\rightarrow r_{+} \ ,
\label{eq:bc}
\end{eqnarray}
where $J_\beta(x)$ is a Bessel function, and has been expanded in the second line for small $r-r_+$.
This solution represents a wave traveling in the radial direction, whose group velocity is given by $v_g=\pm 4\pi T$. Therefore, the solution that is in-going at the horizon is obtained by imposing as a boundary condition the form \eqref{eq:bc} with the ``$-$'' sign. In practice, for numerical calculations we choose a cutoff $\varepsilon$ and impose the boundary condition as
\begin{equation}
R\left(r_+\!+\varepsilon\right) =\varepsilon ^{-i\frac{\omega-\omega_s }{4\pi T}}
\ .
\label{eq:bcepsilon}
\end{equation}
Notice that, in order for the approximated form \eqref{eq:bc} to be valid, we need to discard the $\mu_l^2$ term in \eqref{eqRarbDapp} which implies that our cutoff must satisfy $\varepsilon\ll (\omega-\omega_s)^2/4\pi T \mu_l^2$.

The resulting phase velocity $v_f=-4\pi T \omega_R/(\omega_R-\omega_s)$ is outgoing whenever $\omega_R$ lies in the superradiant region $\omega_R<\omega_s$. This implies that the black hole is radiating monochromatic waves at its horizon. Since the energy flux at the horizon is given by the energy-momentum tensor as $T^{0r}\propto\omega_R^2/v_f(r-r_+)$, it is emitting energy. This ``superradiant'' behavior strongly relies on having a non-vanishing temperature, and a non-vanishing $\omega_s$. This last condition implies, in particular, that the Euclidean continuation is not regular at the horizon. Therefore, superradiance is an out of equilibrium phenomenon.

In such scenario one can imagine that, if the superradiant monochromatic modes are reflected somewhere in the outer region and sent back to the black hole, they could interfere constructively with the emitted waves, resulting in an amplification. This amplification could then accumulate with further reflections, and lead to an instability. Such instability would show up as an exponential temporal growth of the amplitude, {\em i.e.} as a positive imaginary part of the frequency $\omega_I>0$.
To test this, we consider as a boundary condition in the outer region a ``reflecting mirror'', {\em i.e.} a spherical shell at a finite radius $r_m$ at which the scalar vanishes
\begin{equation}
R(r_m)=0\,.
\label{eq:mirror}
\end{equation}

The field starts at the cutoff $r=r_++\varepsilon$ with the value given by \eqref{eq:bcepsilon}, and then oscillates as a function of $r$ as we move towards the mirror, with a wavelength that is determined by the frequency. In order to satisfy \eqref{eq:mirror} we need to complete at least half a wavelength before reaching the mirror, or an integer or half-integer number of them. This results in quantized frequencies.
The upper bound $\omega_R<\omega_s$ on the superradiant frequencies implies a lower bound on the wavelengths. Therefore, in order to fit an integer or half-integer number of wavelenghts, we need to put the mirror at large enough $r_m$. This implies that the superradiant instability is a long wavelength one.

\begin{figure}
\setlength\unitlength{1mm}
\includegraphics[width=110mm]{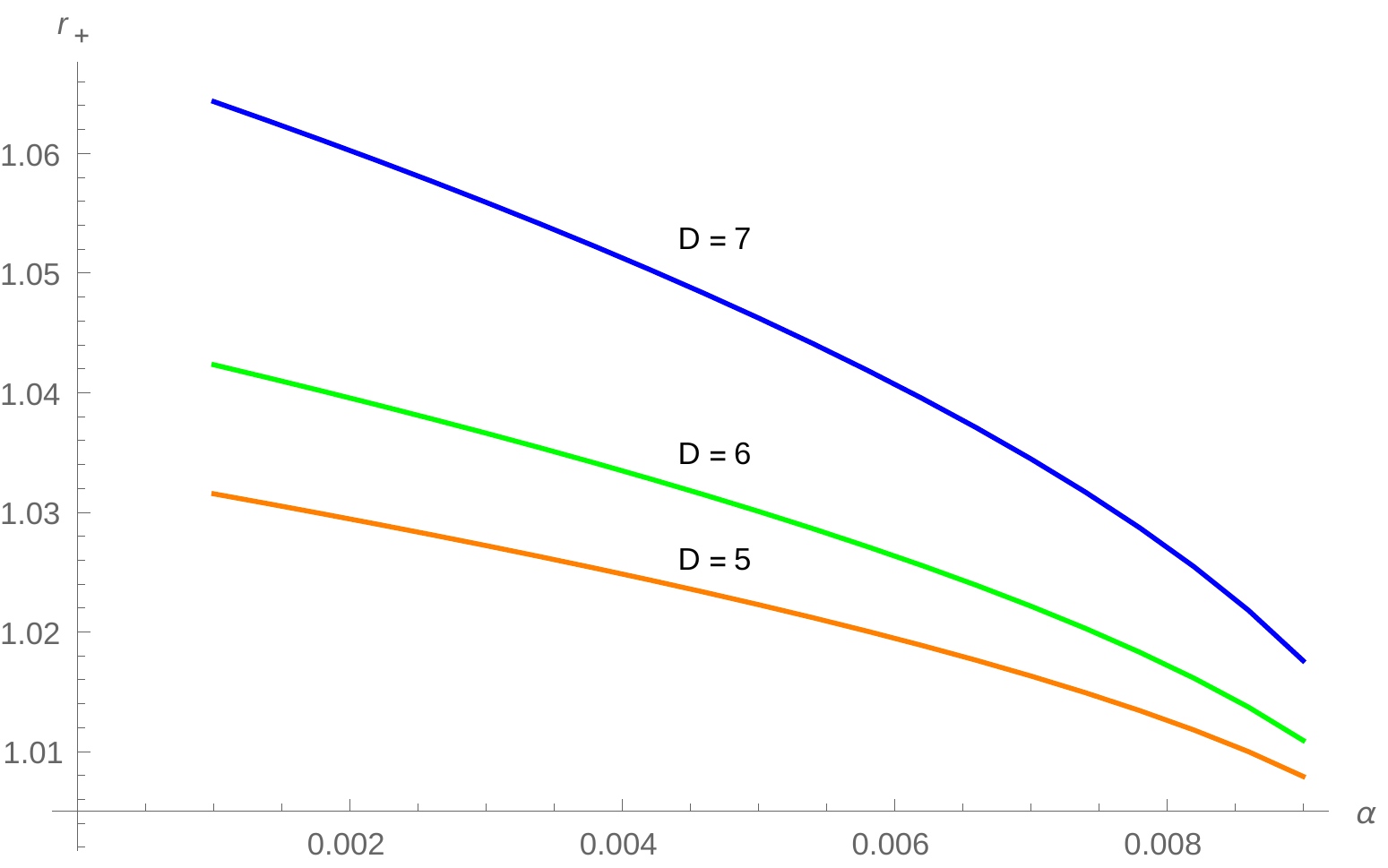}
\caption[FIG:A]{The horizon radius as a function of the Gauss-Bonnet coupling $\alpha$ in $D=5,6,7$ dimensions. We see that the radius decreases as $\alpha$ grows. The plot correspond to $M=1$, $Q=0.99$, $\mu=0.3$,$q=0.6$, $l=1$.}
\label{fig.0}
\end{figure}

In the Gauss-Bonnet black hole, the horizon radius shrinks as a function of the Gauss-Bonnet parameter $\alpha$ (see Fig.\ref{fig.0}). Moreover, for smaller horizon radius the upper bound on the superradiant frequencies $\omega_s=-qA_t(r_+ )$ is larger, implying that the lower bound in wavelenghts is smaller. Both observations add up to the expectation that the instability radius would shrink as $\alpha$ grows. Nevertheless, this shrinking must have a limit, {\em i.e.} a minimum radius of the mirror at which the instability exits. This can be verified as follows: if the mirror is put arbitrarily close to the horizon, the form \eqref{eq:bc} is a good approximation of the solution on the whole range $r_+$ to $r_m$, but it cannot satisfy \eqref{eq:mirror} for any finite value of the frequency. In consequence, as $\alpha$ grows the system becomes unstable at smaller values of $r_m$, but there must be always a finite distance between the horizon and the mirror. In the following section, we explore numerically these hypotheses.

\section{Numerical results}

We integrated numerically the radial equation (\ref{eqRarbD}) for $D=5,6$ and $7$ with boundary conditions \eqref{eq:bcepsilon} and \eqref{eq:mirror}, by using a shooting method implemented in {\tt Mathematica}. We explored the spectrum in order to identify the fundamental frequency, that is the one with the smallest real part (that at the same time has the smallest imaginary part). By moving $r_m$ we were able to identify the value at which the imaginary part of the fundamental frequency becomes positive. At such value of the mirror radius, the system becomes unstable.

\begin{figure}
\setlength\unitlength{1mm}
\vspace{-2.5cm}
\includegraphics[width=110mm]{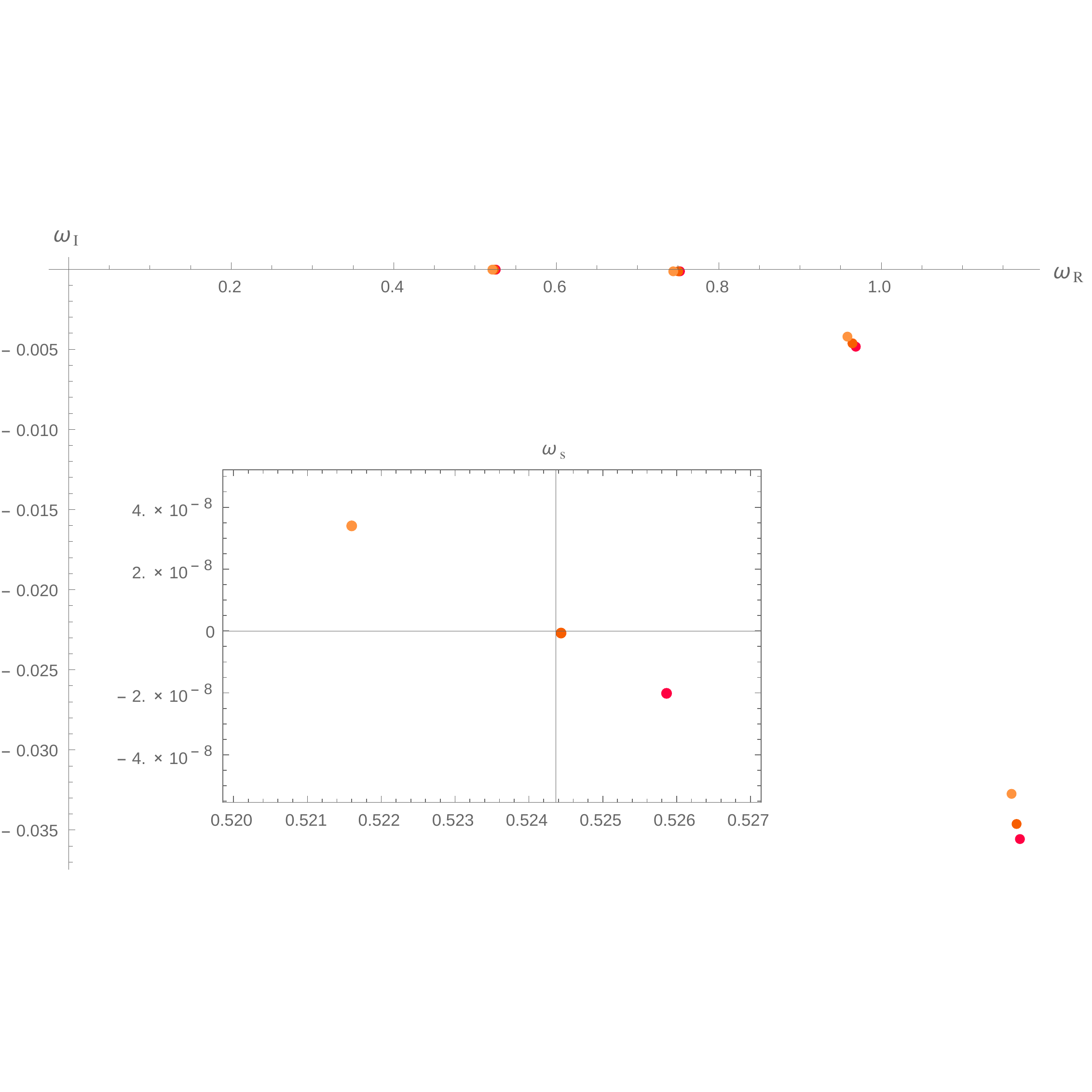}\\
\vspace{-2.5cm}
\caption[FIG:A]{The lowest part of the spectrum in the complex frequency plane in $D=5$ with $\alpha=1/1000$ for three values of the mirror radius $r_m=11.85$ (red), $r_m=11.9$ (orange), $r_m=12.0$ (yellow). As the radius grows, the spectrum approaches the real axes. As can be seen in the inset, the fundamental frequency, {\em i.e.} the one with the smallest real part, is the first in crossing to the upper half plane $\omega_I>0$ at the same time as it enters the superradiant region $\omega_R<\omega_s$. The plot correspond to $M=1$, $Q=0.99$, $\mu=0.3$,$q=0.6$, $l=1$.}
\label{fig.1}
\end{figure}

The spectrum for different values of $r_m$ at fixed $\alpha$ and dimension $D=5$ is shown in Fig.\ref{fig.1}. As mentioned above, the fundamental frequency is the one with the smallest real part $\omega_R$. It can be checked that for $\alpha$ and $r_m$ in an appropriate range, the fundamental mode has a negative imaginary part $\omega_I<0$. Excited states are identified as those with a larger real part, all of them having imaginary parts that lie at more negative values in the imaginary axes, {\em i.e.} they represent more strongly damped modes. As $r_m$ grows, the fundamental frequency  goes into the superradiant region $\omega_R<\omega_s$, and at the same time it enters the upper half complex plane $\omega_I>0$ signaling the instability.

\begin{figure}
\setlength\unitlength{1mm}
\vspace{-1cm}
\hspace{-2cm}
\includegraphics[width=150mm]{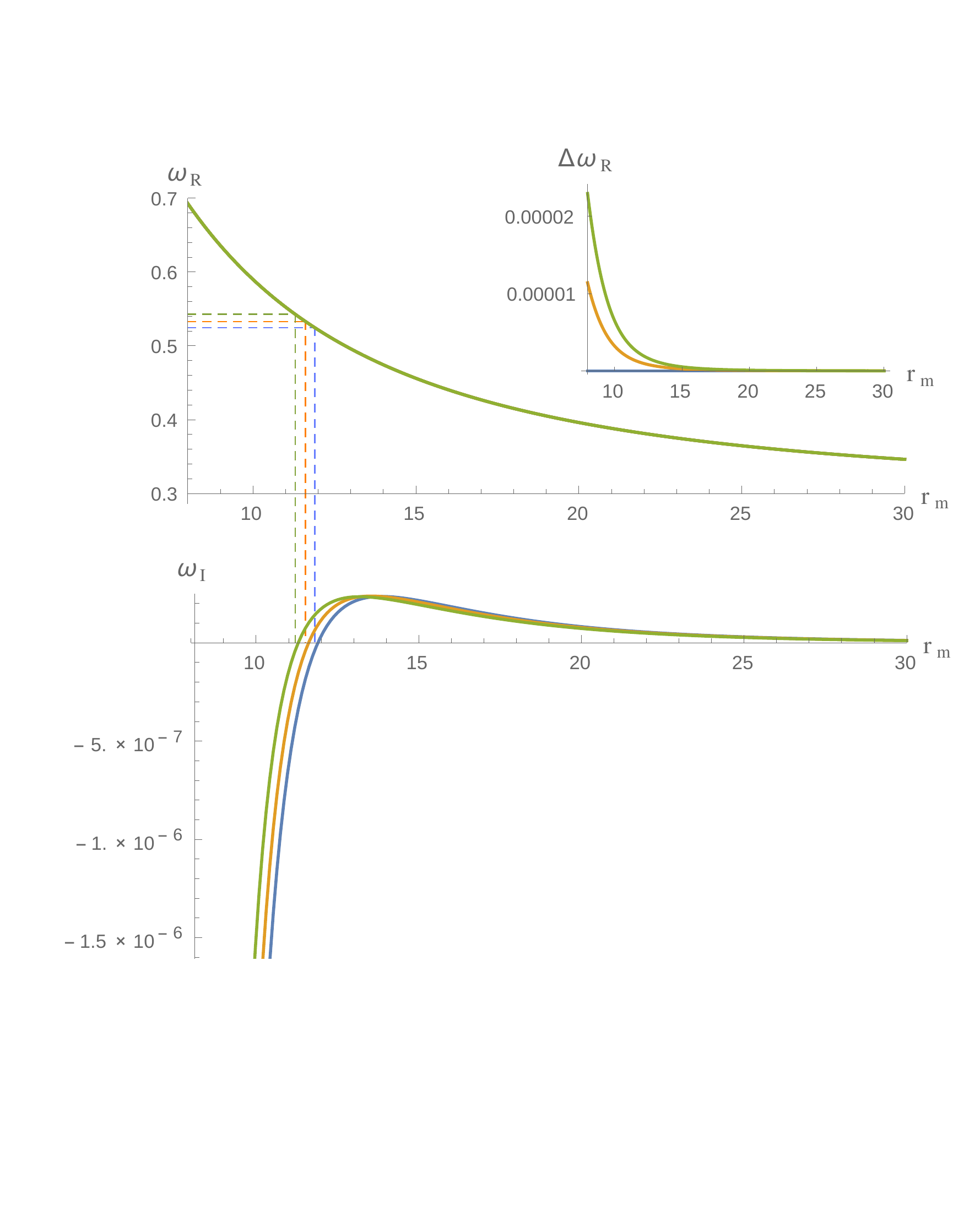}
\vspace{-3.5cm}
\caption[FIG:A]{The real (top) and imaginary (bottom) parts of the frequency as a function of $r_m$ in $D=5$ dimensions for different values of $\alpha=1/1000,3/1000,5/1000$ (blue, orange, green, respectively). To make the different behaviors of $\omega_R$ distinguishable, in the inset we substracted the $\alpha=1/1000$ frequency, plotting $\Delta \omega_R = \omega_R-\omega_R|_{\alpha=1/1000}$.
The dotted lines mark in the vertical axes the value of $\omega_s$ and in the horizontal axes the critical value of $r_m$ at which $\omega_R<\omega_s$ and at the same time $\omega_I>0$ signaling the onset of the instability.
We see that such critical radius decreases as $\alpha$ grows. The plot correspond to $M=1$, $Q=0.99$, $\mu=0.3$,$q=0.6$, $l=1$.
}
\label{fig.2a}
\end{figure}

The real $\omega_R$ and imaginary $\omega_I$ parts of the fundamental frequency as functions of $r_m$ for fixed $\alpha$ are shown in Fig.\ref{fig.2a}. As expected, the smaller values of $r_m$ lead to a stable propagation of the scalar which has $\omega_I<0$ when $\omega_R>\omega_S$. This is consistent with our interpretation that for small $r_m$ there is not enough place to complete a wavelength before reaching the mirror, and then the constructive amplification cannot take place. As $r_m$ grows the fundamental mode enters into the superradiant region $(\omega_R<\omega_s)$ and the system becomes unstable $(\omega_I>0)$ when the constructive amplification occurs. When $r_m$ goes to infinity $\omega_I$ goes to zero, consistently with the stability in the absence of a mirror.

\begin{figure}
\setlength\unitlength{1mm}
\includegraphics[width=110mm]{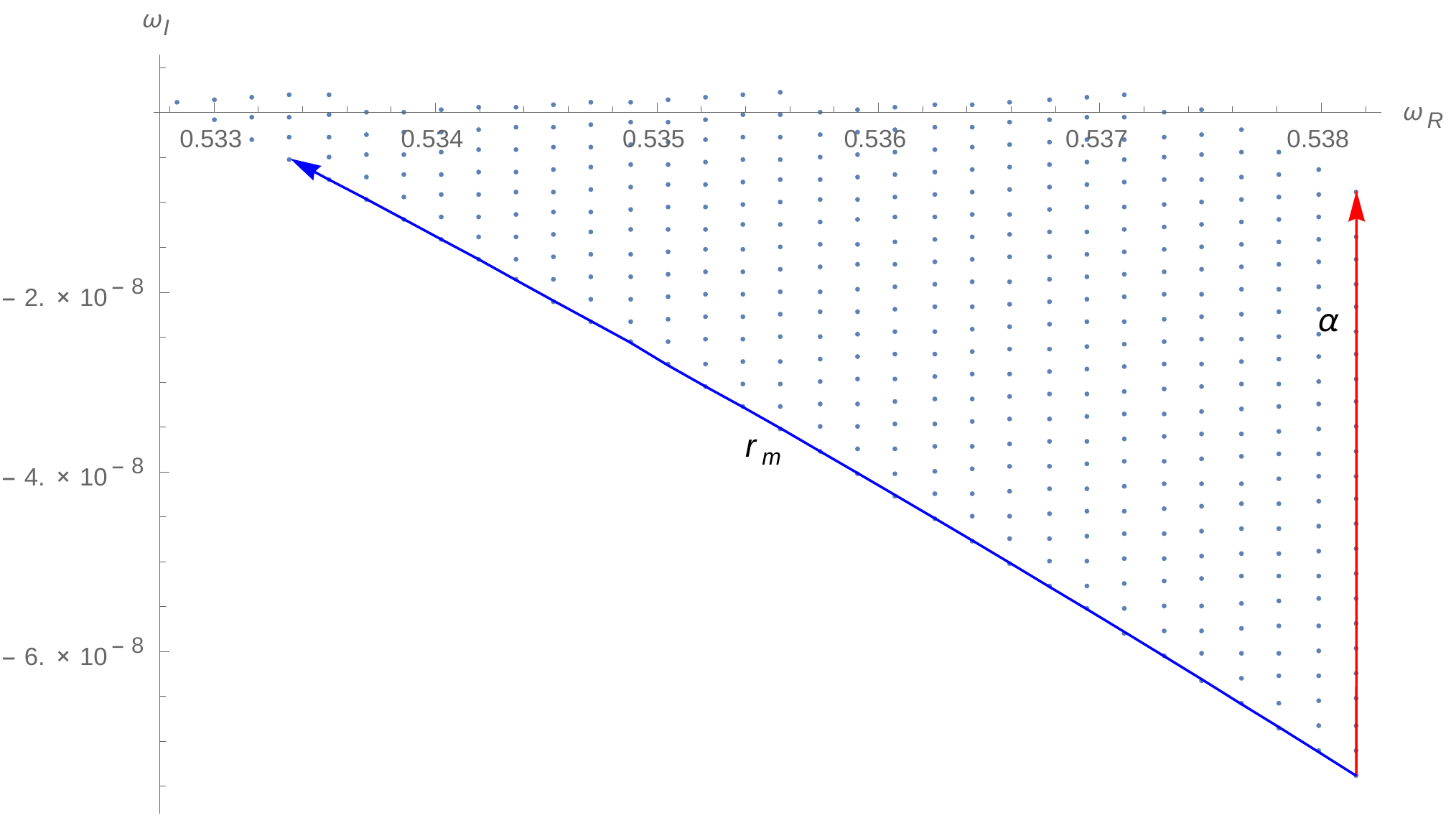}
\caption[FIG:A]{The complex $\omega$ plane with the fundamental frequencies for different values of the Gauss-Bonnet coupling $\alpha$ and the mirror radius $r_m$,  in $D=5$ dimensions. The blue (respectively red) arrow shows the behavior of the frequencies as $r_m$ (respectively $\alpha$) grows. The plot correspond to $M=1$, $Q=0.99$, $\mu=0.3$,$q=0.6$, $l=1$.}
\label{fig.3}
\end{figure}

The fundamental frequency for different values of $\alpha$ and $r_m$ is depicted in the complex plane in Fig.\ref{fig.3}. There, it can be seen that the imaginary part of the frequency approaches the real axes as $\alpha$ grows for fixed $r_m$, and independently as $r_m$ grows for fixed $\alpha$.

\begin{figure}
\setlength\unitlength{1mm}
\includegraphics[width=110mm]{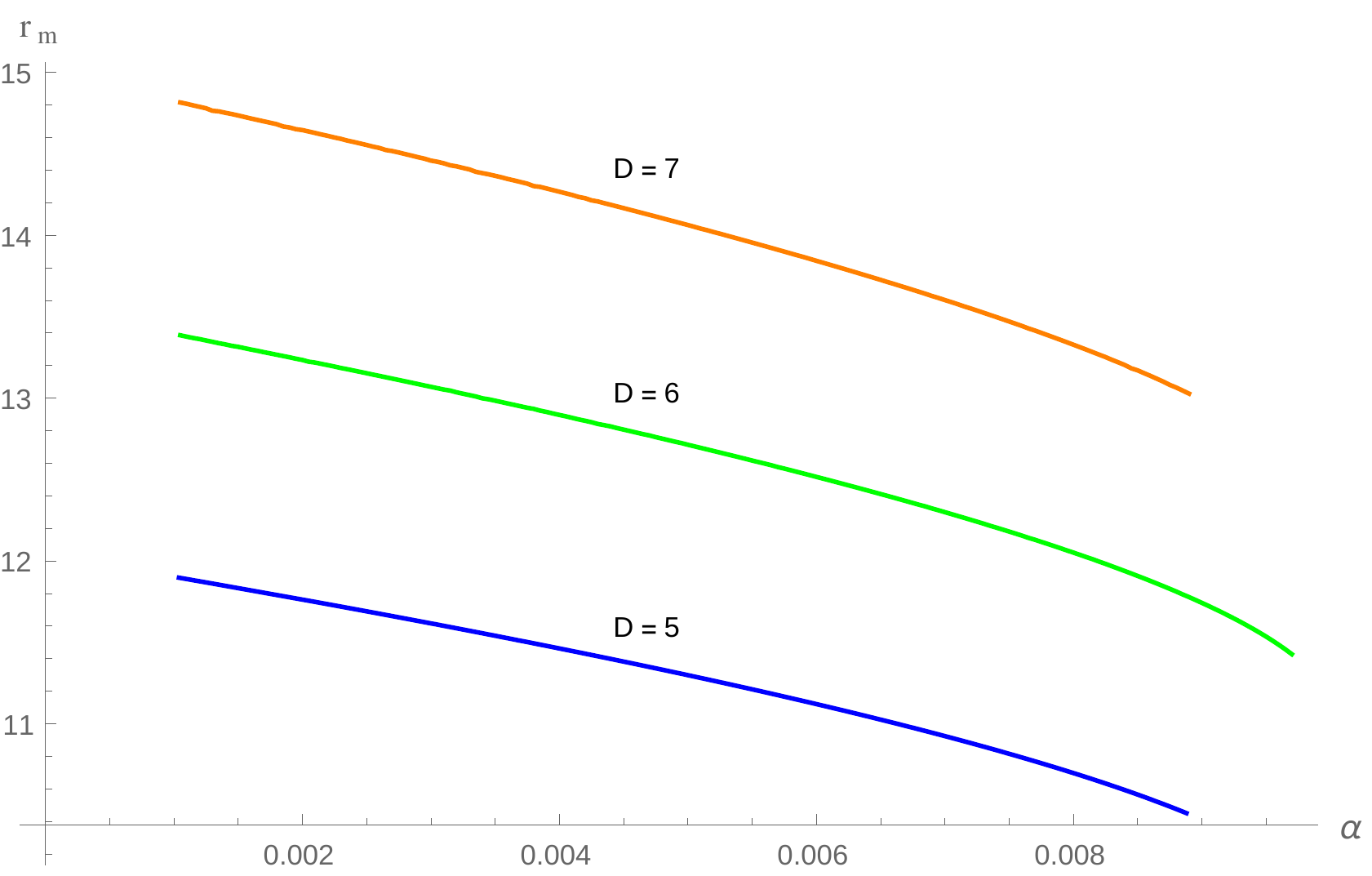}
\caption[FIG:A]{The values of the mirror radius $r_m$ at which the system becomes unstable, as a function of the Gauss-Bonnet coupling $\alpha$, in $D=5,6,7$ dimensions. We see that the critical radius decreases as $\alpha$ grows. The plot correspond to $M=1$, $Q=0.99$, $\mu=0.3$,$q=0.6$, $l=1$.}
\label{fig.4}
\end{figure}

The radius at which the scalar becomes unstable is plotted as a function of $\alpha$ in Fig.\ref{fig.4}. As hypothesized, for larger $\alpha$ the system becomes unstable at a smaller mirror radius. In other words, the instability radius shrinks as $\alpha$ grows, as is happens to the horizon radius.

In $D=5$, for a fixed mass and charge the spherically symmetric black hole solution exists provided $\alpha$ is bounded from above
\footnote{This can also be stated as the fact that for a fixed value of $\alpha$ the black hole exists provided $M$ is greater than a minimum mass. This is similar to what occurs in $2+1$ dimensions with the BTZ black hole \cite{Banados:1992wn}, where there is an energy gap between the maximally symmetric AdS background and the black holes}.
We have checked that, up to the range allowed by our numerical precision, the mirror radius remains at a finite distance of the horizon as $\alpha$ approaches this upper bound. In $D=6, 7$ there is no upper bound for $\alpha$ on the charged solution, and the black holes exist for arbitrarily large values of the Gauss-Bonnet coupling. In these cases, up to the ranges we were able to explore the curves do not intersect. This is again consistent with our hypothesis that there is a minimum distance between the horizon and the mirror for the instability to occur.

\section{Conclusions}
We have shown that electrically charged black holes in Einstein-Gauss-Bonnet gravity suffer from  superradiant instability, triggered by a charged scalar that fulfils Dirichlet boundary conditions at a mirror located in the region of outer communications. As in General Relativity this is a long wavelength instability since it requires the mirror to be located above a minimum critical radius. Such critical value decreases as the Gauss-Bonnet coupling increases, and the mirror approaches the horizon. Nevertheless, we checked that the distance between the mirror and the horizon remains finite in the numerically accessible range. We have explored this phenomenon in dimensions $D=5,6$ and $7$, and the behaviour is similar in all of these cases. 

Our results are consistent with the literature on superradiance in Reissner-Nordstrom solutions, the real and imaginary part of the fundamental frequencies as a function of $r_{m}$, in Fig.\ref{fig.2a}, recovering the corresponding curve of \cite{Herdeiro:2013pia} in the limit of $\alpha \rightarrow 0$.

Stability studies of the black holes in Einstein-Gauss-Bonnet theory are certainly not a new subject. The stability of the Boulware-Deser solution has been explored in detail in the past. The article \cite{Dotti:2004sh} proved the stability of the tensor mode gravitational perturbation for $D\neq6$, while in $D=6$, as well as for the case with a negative curvature horizon in arbitrary dimensions, an unstable family of black holes was found \cite{Dotti:2005sq}-\cite{Beroiz:2007gp}. This analysis was extended to the involved cases of the vector and scalar modes of the gravitational perturbation in \cite{Gleiser:2005ra}, where it was found that the latter generically trigger instabilities. In references \cite{insegblambda1}-\cite{insegblambda3} new instabilities were found, also in the presence of a cosmological constant. The stability of the black holes in Lovelock theories beyond Einstein-Gauss-Bonnet was explored in \cite{Takahashi:2009dz}-\cite{Takahashi:2009xh} finding again a range of parameters for which the black holes turn out to be unstable, which was extended also to the electrically charged case \cite{Takahashi:2011qda}-\cite{Takahashi:2012np}. The quasinormal modes have also been computed in a series of papers. In \cite{Konoplya:2004xx} a thorough analysis of the quasinormal modes for the Boulware-Deser solution in arbitrary dimensions was given. In \cite{Chakrabarti:2005cm} it was shown that the highly damped modes behave differently than their counterpart in General Relativity, and even an analysis for the quasinormal modes in the asymptotically AdS case has recently appear for both the Einstein and non-Einstein branches \cite{Gonzalez:2017gwa}. In all of these studies, the domain where the perturbations evolve is the whole region outside the black hole. The instabilities we have found in the present paper are of a different nature than the instabilities that may appear in the previously mentioned references \footnote{See also references \cite{Li:2014fna}, \cite{Li:2015mqa} for the studies on the superradiant instability of the charged Gibbons-Maeda black hole and for charged hairy black holes in \cite{Gonzalez:2017shu}.}.

A natural question is, as usual, what's the final stage of the instability. In general relativity, the numerical simulations seem to indicate that the final stage of the superradiant instability for the Reissner-Norstrom black hole with a scalar bounded in a cavity, is a hairy charged black hole \cite{Sanchis-Gual:2015lje} even in the presence of a self-interacting potential for the field \cite{Sanchis-Gual:2016tcm}. Such configurations seem to be stable \cite{Dolan:2015dha}. A particularly relevant feature of these processes is the fact that as the gauge coupling $q$ increases the transference of charge from the initial black hole to its surrounding passes from a monotonic behaviour to a very energetic and abrupt process (dubbed ``bosenova" in \cite{Sanchis-Gual:2015lje}). Another interesting topic is of course the study the evolution of charged fields in asymptotically AdS spaces. The causal structure of AdS naturally provides for ``cavity-like" boundary conditions. For the Reissner-Norstrom-AdS solution this has been recently explored in \cite{Bosch:2016vcp}, leading to similar conclusions than the cavity case with an abrupt behavior as $q$ increases. It's natural to expect that such process being highly energetic might receive corrections in the presence of higher curvature terms as the one considered in this paper.

\section{Acknowledgement}
N.E.G. would like to thank Pablo Gonz\' alez-Pisani for helpful discussions. J.O also thanks Carlos Herdeiro for enlightening discussions. This work is partially supported by grants PIP-2008-0396 (Conicet, Argentina) and PID-2013-X648 and PID-2017-X791 (UNLP, Argentina). This work was also partially supported by CONICYT grant PAI80160018 and Newton-Picarte Grants DPI20140053.

\end{document}